\documentclass[a4paper,11pt]{article}
\pdfoutput=1
\usepackage{graphicx}
\usepackage{epsf,amsmath,bbold,amsfonts,stmaryrd}

\usepackage[utf8]{inputenc}
\usepackage{mathrsfs}
\usepackage{appendix}
\usepackage{amssymb}
\usepackage{float}
\usepackage{color}
\usepackage{cite}
\usepackage[colorlinks]{hyperref}
\hypersetup{pageanchor=false}
\usepackage{indentfirst}
\usepackage{url}
\usepackage{float}
\usepackage{caption}
\usepackage[numbers,square,comma,sort&compress,merge]{natbib}

\def\a{\alpha}

\def\mc{\mathcal}

\def\be{\begin{equation}}
\def\ee{\end{equation}}

\def\bea{\begin{eqnarray}}
\def\eea{\end{eqnarray}}

\def\ba{\begin{array}}
\def\ea{\end{array}}

\def\bc{\begin{center}}
\def\ec{\end{center}}

\def\bl{\begin{flushleft}}
\def\el{\end{flushleft}}

\def\br{\begin{flushright}}
\def\er{\end{flushright}}

\def\bi{\begin{itemize}}
\def\ei{\end{itemize}}

\def\bt{\begin{tabular}}
\def\et{\end{tabular}}
\newcommand{\sR}{\mathsf{R}}

\numberwithin{equation}{section}

\begin{document}

\title{\textbf{Einstein-Gauss-Bonnet Black Strings at Large $D$}}
\author{Bin Chen$^{1,2,3}$\footnote{bchen01@pku.edu.cn}~,
Peng-Cheng Li$^{1}$\footnote{wlpch@pku.edu.cn}~,
and
Cheng-Yong Zhang$^{1,3}$\footnote{zhangcy0710@pku.edu.cn}}
\date{}

\maketitle

\vspace{-10mm}

\begin{center}
{\it
$^1$Department of Physics and State Key Laboratory of Nuclear Physics and Technology,\\Peking University, 5 Yiheyuan Road, Beijing 100871, China\\\vspace{1mm}

$^2$Collaborative Innovation Center of Quantum Matter, 5 Yiheyuan Road, Beijing 100871, China\\\vspace{1mm}

$^3$Center for High Energy Physics, Peking University, 5 Yiheyuan Road, Beijing 100871, China
}
\end{center}

\vspace{8mm}

\begin{abstract}
We study  the black string solutions in the Einstein-Gauss-Bonnet(EGB) theory at large $D$. By  using the $1/D$ expansion in the near horizon region we derive the effective equations that describe the dynamics of the EGB black strings. The uniform and non-uniform black strings are obtained as the static solutions of the effective equations. From the perturbation analysis of the effective equations, we find that thin EGB black strings  suffer from the Gregory-Laflamme instablity and the GB term
weakens the instability when the GB coefficient is small, however, when the GB coefficient is large the GB term enhances the instability. Furthermore, we  numerically solve the
effective equations to study the non-linear instability. It turns out that the thin black strings are unstable to developing  inhomogeneities along their length, and at late times they asymptote to the stable non-uniform black strings. The behavior is qualitatively similar to the case in the Einstein gravity. Compared with the black string instability in the Einstein gravity at large D, when the GB coefficient is small the time needed to reach to final state increases, but when the GB coefficient is large the time to reach to final state decreases. Starting from the point of view in which the effective equations can be interpreted as the equations for the dynamical fluid, we evaluate the transport coefficients and find that  the ratio of the shear viscosity and the entropy density agrees with that obtained previously in the membrane paradigm after taking the large $D$ limit.
\end{abstract}
\baselineskip 18pt

\thispagestyle{empty}
\newpage
\section{Introduction}

In higher dimensions $D>4$ the black holes have much richer physics than in four dimensions \cite{EmparanandReal2008,Horowitz2012}. There are black branes whose worldvolumes are flat but look like black holes in the transverse directions. The stability of the black branes  is interesting, as they could be the solitonic solutions in supergravity.  Gregory and Laflamme discovered that the uniform black string (UBS) solution  whose horizon topology is of $S^{D-3}\times S^1$ is unstable if it is thin enough \cite{Gregory1993} (see the review \cite{Harmark0701}). Moreover,  there is a zero mode at the critical point of the Gregory-Laflamme (GL) instability, which indicates the existence of a static branch of non-uniform black string (NUBS) solutions. In the NUBS solutoins, the translation symmetry along the string direction is broken. This non-uniform branch was found perturbatively \cite{Gubser2002,Sorkin2004} and was numerically studied up to $D=15$ \cite{Wiseman2003,Kleihaus2006,Sorkin2006,Headrick2010,Figueras2012,Kalisch:2016fkm,Kalisch:2017bin}. On the other hand, although the GL instability was well studied in perturbation
theory, the fate of the instability in the non-linear regime at late times was not  well understood. In Ref. \cite{Sorkin2004} it was found that above a critical dimension $D^*\simeq 13.5$  the weakly non-uniform black strings have larger horizon areas than the uniform ones. This suggests that the non-uniform black string could be  the possible end-point of the nonlinear evolution of the uniform black string under the GL instability.
Below the critical dimension $D^*$, the numerical simulations \cite{Lehner2010,Lehner2011} give strong evidence that the evolution does not stop at any stable configuration but proceeds  in a self-similar cascade to arbitrary small scales along the string direction and the cosmic censorship maybe violated.

In recent years, it has been found that black holes physics in higher dimensions can be efficiently investigated by using the $1/D$ expansion in the near region of the black hole \cite{Emparan2013}.
The key feature when the spacetime dimension is very large $D\to\infty$ is that, the gravitational field of a black hole is strongly localized near its horizon due to the very large radial gradient ($\partial_r\sim D/r_0$, $r_0$ is the horizon size) of the gravitational potential. As a result, for the decoupled perturbations \cite{Emparan2014} the black hole can be effectively taken as a surface or membrane embedded in the background spacetime \cite{Emparan2015,Suzuki2015,Bhattacharyya1504,Bhattacharyya1511,Dandekar1607}. The membrane is described by the way it is embedded into the background spacetime, and its non-linear dynamics is determined by the effective equations obtained by integrating out the Einstein equations in the radial direction. By solving the effective equations with different embeddings of the membrane, one can construct different black hole solutions and furthermore study their dynamics perturbatively to find the quasinormal modes or determine numerically the end points of their evolutions under the unstable perturbations \cite{Suzuki1506, Emparan1506, Tanabe1510, Tanabe1511, Emparan1602, Tanabe1605, Rozali1607, Dandekar1609, Chen1702}. For example, in \cite{Suzuki1506} the  non-uniform black string solutions were constructed and the phase structure were studied. Furthermore in \cite{Emparan1506} the non-linear evolution of the black string instability was demonstrated and it was shown that at late times the  unstable black strings
in a large enough number of dimension end at stable non-uniform black strings, which proves the conjecture in \cite{Sorkin2004}.
From a broader perspective, the large $D$ (spacetime dimensions) expansion method and the blackfold approach \cite{Emparan0902, Emparan0910} are based on the same philosophy.
The virtue of the large $D$ expansion method is that the spacetime
dimension $D$ provides a natural expansion parameter, independent of the specific solutions.

All these investigations concern the Einstein gravity. In a spacetime dimension $D>4$, the Einstein gravity has a natural generalisation,
the Lovelock higher-curvature gravity of various orders. The most attractive feature of the Lovelock gravity is that its equations of motion are
still the second order differential equations such that the fluctuations around the vacuum do not have ghost-like mode. Among all the Lovelock gravities, the
second-order Lovelock gravity, the so called Einstein-Gauss-Bonnet gravity, is of particular interest. It includes the quadratic terms of the curvature tensors
which appear  as the leading-order correction in the low energy effective action of the heterotic string theory \cite{Zwiebach1985,Boulware1985}.
Although the exact spherically symmetric black hole solution of the EGB gravity theory has been known for quite a long time \cite{Boulware1985,Wheeler1986}, unlike the case of the Einstein gravity, a uniform black string can not be constructed by simply adding a trivial direction to a  spherically symmetric black hole in one lower dimensions\footnote{For the theory with pure Gauss-Bonnet term or a single Lovelock term, it is possible to construct the black string solution analytically \cite{Giribet:2006ec}, just as in pure Einstein gravity. The black string solutions in the Gauss-bonnet gravity in seven dimensions has been discussed in \cite{Giacomini:2015dwa}, and the one in the third order Lovelock gravity in nine dimensions has been studied  in \cite{Giacomini:2016ftc}}. Many efforts on the black string solutions of the EGB gravity theory are based on the numerical analysis in $D=5$ \cite{Kobayashi2005,Suranyi2009} and $D=5\simeq 10$ \cite{Brihaye2010}. These discussions have mostly focused on the construction of the solutions and their thermodynamics, while the issue of the classical non-linear dynamics of the solutions is basically unexplored. The large $D$ expansion method developed recently offers a promising framework to address this issue.

On the other hand, the large $D$ effective equations for the black strings could be interpreted as the equations for the dynamical fluid. From this point of view the transport coefficients was evaluated and found to match well with the fluid/gravity correspondence \cite{Emparan1602}. It would be interesting to extend the study in this aspect to the case of the EGB theory.

The purpose of this work is to study the black string solutions in the EGB theory by using  the large $D$ expansion method. The study of the EGB black holes at large $D$ was initiated in \cite{Chen1511}, for which the focus was on the computation of the quasinormal modes in the large $D$ limit(see also \cite{Kuang1702}).  Furthermore, in \cite{Chen1703} the large $D$ effective theory of EGB black holes was discussed and the instability was studied, the new solution branch at the onset of the instability was analytically constructed as well. Following the footsteps in the Einstein gravity, it should be  possible to construct the black string solutions in the EGB theory and discuss their main features. In section \ref{section2} we solve the EGB equations with proper metric ansatz and obtain the effective equations for the large $D$ EGB black strings. Then we obtain the UBS and NUBS
 as the static solutions of the effective equations, and discuss their thermodynamic quantities. We interpret the effective equations as the ones for the dynamical fluid and then study the properties of the fluid. In section \ref{section3} we investigate the stability of the black string solutions by perturbatively analyzing the effective equations. We find that the black string suffers from the GL instability as well, and then we clarify the effect of the GB term on the instability. Furthermore, we numerically study the evolution of the GL instability in the non-linear regime. We end with a summary and some discussions in section \ref{section4}.

\section{Effective equations}\label{section2}
In this section we consider the large $D$ effective theory for the black strings in the EGB theory. By solving the EGB equations we
derive the effective equations, which contain the information on the mass and
the momentum density of a dynamical black string. In the following for convenience we use $1/n$ as the expansion parameter instead of $1/D$, where
\be
n=D-4.
\ee
\subsection{Set up}
We consider the $D$-dimensional Einstein-Hilbert action supplemented by the GB term:
\be\label{action}
I=\frac{1}{16\pi G}\int d^Dx\sqrt{-g}\biggl(R+\alpha L_{GB}\biggl),
\ee
with
\be
L_{GB}=R_{\mu\nu\lambda\delta}R^{\mu\nu\lambda\delta}-4R_{\mu\nu}R^{\mu\nu}+R^2,
\ee
where $\alpha$ is the GB coefficient, it is positively defined and inversely proportional to the string tension in the heterotic string theory \cite{Boulware1985}. From the action, we obtain the equations of motion for the metric
\be\label{EGBequations}
R_{\mu\nu}-\frac{1}{2}g_{\mu\nu}R+\alpha H_{\mu\nu}=0,
\ee
where
\be
H_{\mu\nu}=-\frac{1}{2}g_{\mu\nu}L_{GB}+2(RR_{\mu\nu}-2R_{\mu\gamma}R^{\gamma}_{\,\,\nu}+2R^{\gamma\delta}R_{\gamma\mu\nu\delta}
+R_{\mu\gamma\delta\lambda}R_{\nu}^{\,\,\gamma\delta\lambda}).
\ee
The simplest solution of EGB theory corresponds to the generalization of the Schwarzschild black holes, $i. e.$ the spherically symmetric EGB black holes.
However, here we are interested in black string solutions which approach asymptotically the $D-1$ dimensional Minkowski space times a circle, $\mathcal{M}^{D-1}\times S^1$. Using the ingoing Eddington-Finkelstein coordinates, the background metric is
\be\label{background}
ds^2=-dv^2+2dvdr+dz^2+r^2 d\Omega_{n+1}^2,
\ee
where $z$ is a coordinate of the compact direction with period $L$. Then we make the metric ansatz as
\be\label{metricansatz}
ds^2=-A dv^2+2(u_v dv+u_z dz) dr-2C_z dzdt+G_{zz}dz^2+r^2 d\Omega_{n+1}^2.
\ee
 The requirement that (\ref{metricansatz}) should asymptotically approach to (\ref{background}) asks that  $A, u_v, G_{zz}\to1$ and $u_z, C_z\to0$ at $r\to\infty$. So in this case the embedding of the membrane is different from the one in \cite{Chen1703}, in which the background is a (A)dS spacetime. The functions in the metric generally depend on $(v,r,z)$. Note that the form of the metric ansztz is the same as the one for the Einstein gravity \cite{Emparan1506}, similarly the large $D$ EGB black holes have the same metric ansatz form as the black holes in the Einstein gravity \cite{Chen1703}.

In order to do the $1/n$ expansion properly we need to specify the large $D$ behaviors of the metric functions. Due to fact that we do not have the closed form of the UBS solutoin in the EGB theory as a reference, the large $D$ scalings of the metric functions are unclear at present. However, the  discussion in the Einstein gravity \cite{Suzuki1506,Emparan1506} may provide us with some useful indications. As we know that the zero-mode  wavenumber of the GL instability of the black string in the Einstein gravity scales like $k_{GL}\simeq\sqrt{n}/r_0$ \cite{Emparan2013, Asnin0706}, this indicates one should rescale $z\to z/\sqrt{n}$ to capture the unstable fluctuations. Here we  assume that this works for the case in the EGB theory as well. In the following we will justify  this assumption by the explicit computation of $k_{GL}$. In addition we consider a small velocity $\mc O(1/\sqrt{n})$ along the string direction. Therefore the large $n$ scalings of the metric functions are respectively
\be\label{largeDscalings}
A=\mc O(1), \quad u_v=\mc O(1) \quad u_z=\mc O(n^{-1}),\quad C_z=\mc O(n^{-1}), \quad G_{zz}=\frac{1}{n}\Big(1+\mc O(n^{-1})\Big).
\ee
By a gauge choice we can set $u_z=0$. Note that in the case of Einstein gravity, inspired by the metric form of a UBS boosted along the string direction, the leading order part of $G_{zz}$ is completely determined \cite{Emparan1506}. In the EGB theory,  as we do not have the closed form of the UBS solution, here we make an assumption that  the leading order part of $G_{zz}$ is the same as the one in the case of Einstein gravity, by which the leading order EGB equations can be consistently solved as will be shown in the following.

At large $D$ the radial gradient becomes dominant, that is $\partial_r= \mc O(n)$, $\partial_v= \mc O(1)$, $\partial_z =\mc O(1)$, so in the near region of the black hole it is better to use a new radial coordinate $\sR$ defined by
\be
\sR=\Big(\frac{r}{r_0}\Big)^n,
\ee
such that $\partial_\sR= \mc O(1)$, where $r_0$ is a horizon length scale which can be set to be unit $r_0=1$.
To solve the EGB equations we need to specify boundary conditions at large $\sR$, they are given by \cite{Emparan2014}
\be\label{bdycondition}
A=1+\mc O(\sR^{-1}),\quad C_z=\mc O(\sR^{-1}),\quad G_{zz}=\frac{1}{n}\Big(1+\mc O(\sR^{-1})\Big).
\ee
On the other hand the solutions have to be regular at the horizon.

In the following as in \cite{Chen1703} we use $\tilde{\alpha}$ instead of $\alpha$ in doing the $1/n$ expansion, with
\be
\tilde{\alpha}=\alpha n(n+1).
\ee
The reason for this choice can be seen by observing the large $D$ limit of the metric of the spherically symmetric EGB black holes. The metric is  given by
\cite{Boulware1985, Wheeler1986}
\be
ds^2=-f(r)dt^2+f^{-1}(r)dr^2+r^2d\Omega^2_{n+2},
\ee
where
\be
f(r)=1+\frac{r^2}{2\tilde{\alpha}}\biggl(1-\sqrt{1+\frac{64\pi G\tilde{\alpha}M}{(n+2)\Omega_{n+2}r^{n+3}}}\biggl).
\ee
In the metric function $f(r)$, $M$ is the mass of the black hole which in terms of the horizon radius $r_H$ can be expressed as
\be
M=\frac{(n+2)\Omega_{n+2}r_H^{n+1}}{16\pi G}\Big(1+\frac{\tilde{\alpha}}{r_H^2}\Big).
\ee
In terms of $\sR$, at leading order of $1/n$ expansion, $f(r)$ becomes
\be
f(\sR)=1+\frac{1}{2\tilde{\alpha}}\biggl(1-\sqrt{1+\frac{4\tilde{\alpha}}{\sR}r_H^{n+1}\Big(1+\frac{\tilde{\alpha}}{r_H^2}\Big)}\biggl).
\ee
From the  above we can see that the solution is reduced to the one in the Einstein gravity if $\tilde{\alpha}$  is very small, i.e.  $\tilde{\alpha}\to0$. However, if $\tilde{\alpha}$ is very large, e.g. $\tilde{\alpha}=\mc O(n^2)$,  $f(\sR)$ becomes
\be
f(\sR)=1-\frac{1}{\sqrt{\sR}},
\ee
at the leading order of the $1/n$ expansion, where we take $r_H=1$. So in this case the solution cannot smoothly connect to the one in the Einstein gravity.

In this paper we focus on the case $\tilde{\alpha}=\mc O(1)$ when doing the $1/n$ expansion, which can clearly  capture the effect of the GB term. The results in this case can be smoothly extrapolated  to the ones in the Einstein gravity by taking the limit $\tilde{\alpha}\to 0$. On the other hand, we can consider the case with a large $\tilde{\alpha}$. In the appendix \ref{appendixA}, we study the case $\tilde{\alpha}=\mc O(n^2)$ (the case $\tilde{\alpha}=\mc O(n)$ or $\tilde{\alpha}=\mc O(n^3)$ is similar). By imposing a different boundary conditions,
we find that  the leading order results in the large $\tilde{\alpha}$ case can be related to the ones in the case  $\tilde{\alpha}=\mc O(1)$ after taking
the appropriate  large  $D$ scalings, and the effective equations are the same as the ones in the Einstein gravity.
It turns out that the scaling of the GL threshold mode is the same as that in the Einstein gravity and the assumption for the metric components
(\ref{largeDscalings}) works in that case as well.

\subsection{Effective equations}

At the leading order of the $1/n$ expansion, the EGB equations (\ref{EGBequations}) only contain $\sR$-derivatives so they can be solved by performing $\sR$-integrations. Then after
imposing the boundary conditions the leading order solutions are obtained as
\be\label{LOAm}
A=1+\frac{1}{2\tilde{\alpha}}\Bigg( 1-\sqrt{1+\frac{4\tilde{\alpha}\, m(v,z)}{\sR}}\Bigg),\quad u_v=1
\ee
\be\label{LOCzGzz}
C_{z}=\frac{p_z(v,z)}{2\tilde{\alpha}\,m(v,z)}\Bigg(-1+\sqrt{1+\frac{4\tilde{\alpha}\, m(v,z)}{\sR}}\Bigg),\quad G_{zz}=\frac{1}{n}\Big(1+\frac{G_0(v,z)}{n}\Big),
\ee
where
\bea\label{G0}
G_0&=&\frac{(m^2-p_z \partial_zm+m\partial_z p_z)}{(1+\tilde{\alpha})m^2}\Bigg( 2\arctan \sqrt{1+\frac{4\tilde{\alpha}m}{\sR}}-\frac{\pi}{2}+\frac{\ln \Big(1+\frac{2\tilde{\alpha}m}{\sR}\Big)}{1+2\tilde{\alpha}}-2\ln\frac{1+\sqrt{1+\frac{4\tilde{\alpha}m}{\sR}}}{2} \Bigg)\nonumber\\
&&+\frac{p_z^2}{2\tilde{\alpha}m^2}\sqrt{1+\frac{4\tilde{\alpha}m}{\sR}}
-\frac{p_z^2}{2\tilde{\alpha}m^2}.
\eea
Note that $1/n^2$ term in $G_{zz}$ is obtained at the next-to-leading order in the $1/n$ expansion of the EGB equations. It has to be included since it also appears in the EGB equations at the leading order of the $1/n$ expansion.  The expression of $G_0$ seems a little complicated, in the limit $\tilde{\alpha}\to0$ it reproduces the simple expression of the one in the Einstein gravity \cite{Emparan1506}.

In the above expressions, $m(v,z)$ and $p_z(v,z)$ are the integration functions of $\sR$-integrations of the EGB equations. Physically they can be viewed as the mass and momentum density of the solution. We can see that the horizon of this dynamical black string solution is at
\be\label{horizon}
\sR_H=\frac{m(v,z)}{1+\tilde{\alpha}}.
\ee
At the next-to-leading order of the $1/n$ expansion, we are able to find non-trivial conditions on which $m(v,z)$ and $p_z(v,z)$ should satisfy.
The non-trivial conditions are just the effective equations for the large $D$ EGB black strings. These equations are
\be\label{effeq1}
\partial_v m-\partial_z^2 m=-\partial_z p_z,
\ee
\bea\label{effeq2}
&&\partial_v p_z-\partial_z^2 p_z-\partial_zm+\partial_z \Big(\frac{p_z^2}{m}\Big)\nonumber\\
&&+\frac{2\tilde{\alpha}}{(1+\tilde{\alpha})(1+2\tilde{\alpha})}\Big(\partial_z^2 p_z-\frac{p_z\partial_z^2 m}{m}
-\frac{\partial_z m \partial_z p_z}{m}+\frac{p_z(\partial_zm)^2}{m^2}+\partial_zm\Big)=0.
\eea
In the limit $\tilde{\alpha}\to 0$, the second equation becomes
\be
\partial_v p_z-\partial_z^2 p=\partial_zm-\partial_z \Big(\frac{p_z^2}{m}\Big),
\ee
which reproduces the one in the Einstein gravity \cite{Emparan1506}.

From these equations we can obtain the static solutions, including the uniform and non-uniform black string solutions. These equations can describe  non-linear dynamical evolution of the fluctuations of the black strings.

\subsection{Static string solutions}

In order to find the static solutions, we may further assume that $m(v,z)=m(z)$ and $p_z(v,z)=p_z(z)$.  It is straightforward to find static solutions from the effective equations. From (\ref{effeq1}) we have
\be\label{pz}
p_z(z)=m'(z)+p_0,
\ee
here $p_0$ is an integration constant describing the momentum along the $z$ direction, we can set it to zero.

\paragraph{Uniform black strings} In this case the solution is translationally invariant along the $z$ direction. Then from (\ref{pz})  we obtain
\be\label{UBS}
m(z)=m_0,\quad  p_z(z)=0,
\ee
where $m_0$ is an integration constant and is related to the horizon radius by (\ref{horizon}). We can set $m_0=1+\tilde{\alpha}$ such that for the uniform black strings
the horizon radius is unit. With $m(z)$ and $p_z(z)$, the uniform EGB black string is obtained analytically. If the $1/n^2$ term in $G_{zz}$ is not taken into account then
in the large $D$ limit, the leading order solution is obtained by  adding a trivial direction to a spherically symmetric black hole in one lower dimensions. However, this is not a solution to the EGB theory. This suggests that we should include the $1/n^2$ corrections. However even with the $1/n^2$ correction,  after plugging (\ref{UBS}) into (\ref{G0}) we still can not get a simple expression for $G_0$, in contrast to the uniform black string in the Einstein gravity which has $G_0=0$.

\paragraph{Non-uniform black strings}
From (\ref{horizon}) we have
\be
r^n=\frac{m(z)}{1+\tilde{\alpha}}\equiv \tilde{m}(z).
\ee
When $n$ is large, this is
\be
r=1+\frac{\ln\tilde{m}(z)}{n},
\ee
so  $\ln(\tilde{m}(z))$  describes a small $\mc O(1/n)$ deformation of the uniform surface at $r=1$. In this case  the amplitude of the non-uniformities along the string is of order  $\mc O(1/n)$.

Define
\be
\mathcal{P}(z)=\ln\tilde{m}(z),
\ee
then (\ref{effeq2}) becomes
\be\label{Pz}
\mathcal{P}'''(z)+\mathcal{P}'(z)\mathcal{P}''(z)+\mathcal{P}'(z)=0.
\ee
As a consequence, the equation for $\mathcal{P}(z)$ is identical to that in the Einstein gravity. Firstly, it is easy to consider the near uniform solution, $i.e.$ small $\mathcal{P}(z)$. In this case  $\mathcal{P}(z)$ is obtained as
\be\label{smalldeformation}
\mathcal{P}(z)=\epsilon \cos(z),
\ee
where we assumed the reflection symmetry at $z=0$. We can find that this slightly non-uniform black strings in the EGB theory have the same leading behavior of the GL threshold wavenumber as the one in the Einstein gravity, $i.e.$ $k_{GL}\simeq \sqrt{n}/r_0$, which immediately shows that the assumption (\ref{largeDscalings}) is reasonable.

The equation (\ref{Pz}) can be integrated twice to obtain
\be\label{NBS}
\frac{1}{2}\mathcal{P}'^2+U(\mathcal{P})=E,
\ee
with
\be
U(\mathcal{P})=\mathcal{P}+m_0\, e^{-\mathcal{P}}.
\ee
Here $E$ and $m_0$ are integration constants. Hence the above equation can be regarded as the classical one-dimensional motion of a particle, with the position $\mathcal{P}$, the time $z$, the potential $U$, and the energy $E$. In this case it is easy to obtain analytical approximations and numerical solutions. For example, for large deformations which
correspond to large $E$, $E\gg1$, $\mathcal{P}$ has a parabolic profile.


\subsection{Thermodynamics}
The thermodynamic properties of these solution are determined by the asymptotic charges  including the mass and tension, and the quantities on the horizon including the temperature and the entropy.
In the static case $\partial_v$ becomes a Killing vector, as a result the surface gravity $\kappa$ can be expressed as
\bea
\kappa&=&\frac{n}{2}\sR \partial_\sR A\Big|_{\sR_H}\\
&=& \frac{n}{2}\frac{1+\tilde{\alpha}}{1+2\tilde{\alpha}},
\eea
which is clearly constant. Then the temperature is given by
\be
T_H=\frac{\kappa}{2\pi}=\frac{n}{4\pi}\frac{1+\tilde{\alpha}}{1+2\tilde{\alpha}}.
\ee
The entropy of a black object in EGB theory can be written as an integral over the event horizon via the Wald formula \cite{Wald1993}
\be
S=\frac{1}{4G}\int_{\Sigma_h}d^{n+2}x\sqrt{h}(1+2\alpha \tilde{R}),
\ee
where $h$ is the determinant of the induced metric on the horizon and  $\tilde{R}$ is the event horizon curvature. For the static solutions we have
\be
S=\frac{\Omega_{n+1}}{4G}\frac{1+2\tilde{\alpha}}{1+\tilde{\alpha}}\int m(z) dz.
\ee
Then
\be
T_H S=\frac{\Omega_{n+1}n }{16\pi G}\int m(z) dz,
\ee
therefore at leading order the effect of the GB term disappears.
The physical quantities of a configuration that can be measured asymptotically  in the transverse space are the mass $M$ and the  tension $\mathcal{T}$ in
 the direction of the circle. Similar to the Einstein gravity, these quantities are defined in terms of two constants $c_v$, $c_z$ which appear in the asymptotics of the metric functions
 \be
 g_{vv}\simeq -1+\frac{c_v}{r^{D-4}},\quad g_{zz}\simeq 1+\frac{c_z}{r^{D-4}}.
 \ee
 Then the ADM formula determines the mass and tension as \cite{Harmark2004}
 \be\label{mass}
\mathcal{M}=\frac{\Omega_{D-3}L}{16\pi G}[(D-3)c_v-c_z],\quad \mathcal{T}=\frac{\Omega_{D-3}}{16\pi G}[c_v-(D-3)c_z],
 \ee
 where $\Omega_{D-3}$ is the volume of $S^{D-3}$.

For the non-uniform black strings,
\be
c_v=\frac{1 }{L}\int m(z)dz,\quad c_z=\mc O(1/n).
\ee
as we have
\be\label{masstension}
\mathcal{M}=\frac{\Omega_{n+1} (n+1) }{16\pi G}\int m(z)dz,\quad \mathcal{T}=\frac{\Omega_{n+1} }{16\pi G}\cdot\mc O(1).
\ee
It is easy to see the Smarr's formula is satisfied trivially at leading order of $1/n$ expansion
\be
T_H S=\mathcal{M}.
\ee
Since the tension of the black string is of $\mc O(1)$, both $\mathcal{M}$ and $T_H S$ are of $\mc O(n)$, so $\mathcal{T}$ does not contribute to the
  Smarr formula at the leading order of the $1/n$ expansion. From (\ref{masstension}) we see that like the case in the Einstein gravity \cite{Emparan2013},  the tension of a EGB black string is  small compared with its mass at large $D$. The effect of the GB term is reflected in the $1/n$ corrections to the  mass and the  tension. A direct application of this fact is that we might be able to construct a EGB black ring solution
 by  bending and rotating the EGB black string. Intuitively one can imagine that a black ring as a rotating bent string such that the centrifugal force balances the string tension. The analysis by the blackfold method found \cite{Emparan0708} that the horizon angular velocity of the $D$ dimensional thin black ring is of $\mc O(1/\sqrt{D})$. At large $D$, since the string tension is small, the horizon angular velocity of a large $D$ black string should be small as well. Using the large $D$ effective theory, a neutral black ring solution was constructed analytically by solving the effective equations for slowly rotating black holes \cite{Tanabe1510}, then the work was extended to the charged case in the Einstein-Maxwell theory at
 large $D$ \cite{Chen1702}. Now that the  EGB black has a small tension, we expect to obtain a EGB black ring by applying the large $D$ effective theory to the slowly rotating black holes in EGB theory \cite{Chen1707}.

\subsection{Dynamical fluid}

In Ref. \cite{Emparan1602}, it was showed that the effective equations for the Einstein black strings (branes) could be interpreted as the equations for the dynamical fluid. In this subsection, following this idea we would study the properties of the dynamical fluid  in the EGB membrane. Firstly we define the velocity variable as
\be
p_z=m u_z+\partial_z m,
\ee
in terms of which, the first effective  equation  (\ref{effeq1}) becomes the continuity equations for the mass density,
\be
\partial_v m +\partial_z(m u_z)=0,
\ee
as the one in the case of the Einstein gravity. Moreover, in terms of $u_z$, the  effective equation (\ref{effeq2}) becomes
\bea\label{momentumstresseq}
&&\partial_v(m u_z)+\partial_z(m u_z^2+\tau_{zz})=0,
\eea
where
\bea
\tau_{zz}&=&-\frac{1+\tilde{\alpha}+2\tilde{\alpha}^2}{(1+\tilde{\alpha})(1+2\tilde{\alpha})} m-\frac{2(1+2\tilde{\alpha}+2\tilde{\alpha}^2)}{(1+\tilde{\alpha})(1+2\tilde{\alpha})}m \partial_z u_z-\frac{1+\tilde{\alpha}+2\tilde{\alpha}^2}{(1+\tilde{\alpha})(1+2\tilde{\alpha})}m\partial_z^2 \ln m.
\eea
We may view these as the equations for a non-relativistic, compressible fluid with the mass density $m$, the velocity $u_z$ and the stress
tensor $\tau_{zz}$. In the hydrodynamic gradient expansions, the  third part in (\ref{momentumstresseq}) can be neglected if we keep only the leading order of the spatial gradients  of the variables. Compare with the standard  equations for the fluid we can identify the pressure
\be
P=-\frac{1+\tilde{\alpha}+2\tilde{\alpha}^2}{(1+\tilde{\alpha})(1+2\tilde{\alpha})} m,
\ee
 which reads the equation of state of the black string. The speed of sound of long-wavelength perturbations is then
\be\label{soundspeed}
c_s\equiv\sqrt{\frac{\partial P}{\partial m}}= \sqrt{-\frac{1+\tilde{\alpha}+2\tilde{\alpha}^2}{(1+\tilde{\alpha})(1+2\tilde{\alpha})}}.
\ee
Note that since the physical mass density is $\rho= n m$ (see from (\ref{masstension})), the speed of sound is as small as $\mc O(1/\sqrt{n})$, which also justifies the assumption that the velocity  along the string direction is of $\mc O(1/\sqrt{n})$. A negative pressure and therefore an imaginary $c_s$ for the EGB black strings gives rise to the GL instability, like the case of the Einstein gravity \cite{Emparan0910, Camps1003}. 

From the stress tensor of the fluid, we can read  shear and bulk viscosities  be
\be
\eta=\frac{1+2\tilde{\alpha}+2\tilde{\alpha}^2}{(1+\tilde{\alpha})(1+2\tilde{\alpha})}m,\quad \zeta=2\eta.
\ee
In addition to   the energy density $m$, we have other local thermodynamical variables, the entropy density $s$, and the reduced temperature ($T_H= n T$)
\be
s=\frac{1+2\tilde{\alpha}}{1+\tilde{\alpha}}4\pi m,\quad T=\frac{1}{4\pi}\frac{1+\tilde{\alpha}}{1+2\tilde{\alpha}},
\ee
where we have used the unit $16\pi G=\Omega_{n+1}$. Then besides the generic thermodynamics equations $Ts=m$ and
$ dm=Tds$, we have the ratio of the shear viscosity and the entropy density
\be\label{etas}
\frac{\eta}{s}=\frac{1+2\tilde{\alpha}+2\tilde{\alpha}^2}{(1+2\tilde{\alpha})^2}\frac{1}{4\pi}.
\ee

In \cite{Kovtun0309}, it was  conjectured that through the AdS/CFT correspondence  for all known physical systems
$\eta/s\geq1/4\pi$, which is called the KSS bound. It was found that in the theories dual to the EGB gravity this bound was violated \cite{Brigante0712}. Moreover,  the membrane paradigm \cite{Thorne, Damour} for the EGB gravity in $D\geq5$ has been studied in \cite{Jacobson1107}, in which
the transport coefficient for the black holes with constant curvature horizons and negative or zero
cosmological constant was derived. For the asymptotical flat spacetime, the ratio of the shear viscosity and the entropy density is of the following form in our convention
\be
\frac{\eta}{s}=\frac{1}{4\pi}\frac{1+2\tilde \alpha \frac{D-5}{D-3}(1+\tilde \a)}{(1+2\tilde \a)(1+2\tilde \a\frac{D-2}{D-4})}.
\ee
At the large $D$ limit, this is exactly the same as the relation (\ref{etas}) we obtained before. Obviously, it violates the KSS bound for any $\tilde{\alpha}>0$.

 \section{Instability and non-linear evolution}\label{section3}
In this section, we study the instability of the black string solutions obtained in the last section. For the black string, there could be Gregory-Laflamme instability. This will be discussed in the following subsection. Moreover we will study the non-linear dynamical evolution of the unstable string.

\subsection{Quasinormal modes: Gregory-Laflamme instability}\label{QNM}
\begin{figure}[t]
 \begin{center}
  \includegraphics[width=65mm,angle=0]{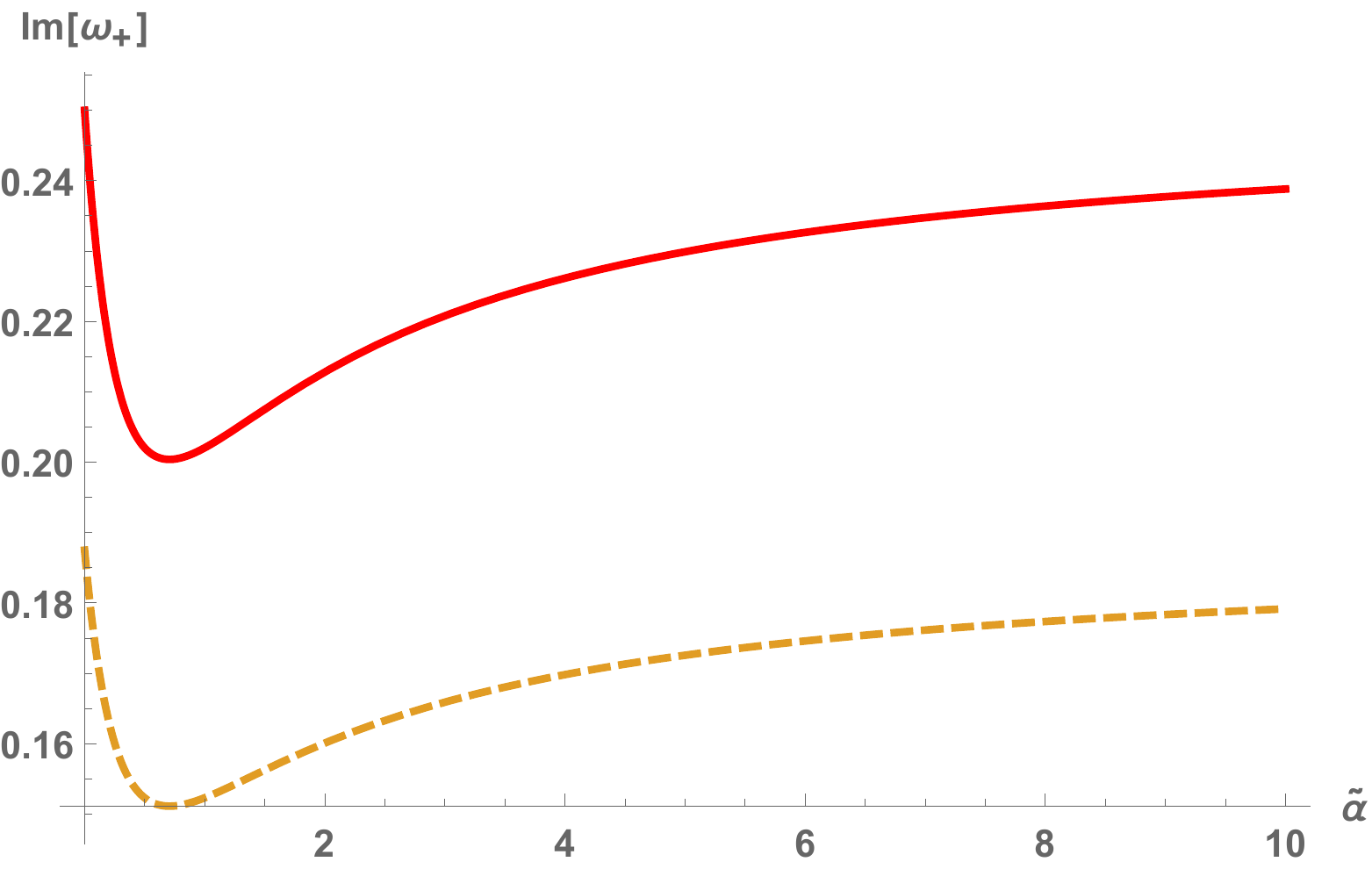}
 \end{center}
 \vspace{-5mm}
 \caption{Dependence of Im[$\omega_+$] of the EGB black string on the GB coefficient $\tilde{\alpha}$. Here the solid line corresponds to $k=1/2$ and the dashed line corresponds to
 $k=1/4$.}
 \label{figfrequency}
\end{figure}

Consider a small perturbations around the static uniform black string (\ref{UBS}), with momentum $k$ aligned along the string direction
\bea
m&=&m_0+\delta m\, e^{-i\omega v+ikz},\\
p_z&=&\delta p_z\, e^{-i\omega v+ikz}.
\eea
The linearized equations of motion of the perturbations could be obtained from the effective equation. After imposing appropriate boundary conditions on the fluctuations at the horizon and at asymptotic region, we may
 obtain the frequencies of the quasinormal modes. For the scalar-type  gravitational perturbations, which are most essential to the stability, we find their frequencies
\be\label{frequency}
\omega_{\pm}=-ik^2\frac{1+2\tilde{\alpha}+2\tilde{\alpha}^2}{(1+\tilde{\alpha})(1+2\tilde{\alpha})}\pm\frac{i k\sqrt{1+4\tilde{\alpha}+(7+k^2)\tilde{\alpha}^2+8\tilde{\alpha}^3+4\tilde{\alpha}^4}}{(1+\tilde{\alpha})(1+2\tilde{\alpha})}.
\ee
To develop an instability the frequency must have a positive imaginary part, $i.e.$ Im[$\omega_+$]$>0$, this requires
\be
k<k_{GL}=1.
\ee
This is the GL instability at large $D$. The threshold wavenumber $k_{GL}$ is the same as the one in the Einstein gravity. In other words, the threshold wavenumber $k_{GL}$ is independent of the GB term.   From (\ref{frequency}) we can easily find that the frequency $\omega_+$ is bounded below and has a minimum at $\tilde{\alpha}=1/\sqrt{2}$ as shown in Fig. \ref{figfrequency}. From Fig. \ref{figfrequency} we can see the effect of the GB term on the GL instability, when
$\tilde{\alpha}<1/\sqrt{2}$ the presence of the GB term makes the instability weaker, which is similar to the role played by the electric charge. In contrast, as $\tilde{\alpha}>1/\sqrt{2}$  the presence of the GB term enhances the instability. From (\ref{frequency}) we can see another feature of the GL instability of EGB black strings, in the limit  $\tilde{\alpha}\to \infty$, $\omega_{\pm}$ becomes
\be
\omega_{\pm}=-ik(k\pm 1),
\ee
the result is identical to the case $\tilde{\alpha}=0$ \cite{Emparan2013} as shown in Fig. \ref{figfrequency}, which is consistent with the analysis of the large $\tilde{\alpha}$ case in the appendix \ref{appendixA}.

\subsection{Non-linear evolution}

 Using the effective equations (\ref{effeq1}) and (\ref{effeq2}) we can numerically  study the non-linear evolution of the black strings. We parametrize the
periodicity $L$ of the string  direction $z$ in terms of a wavenumber $k_L$ as
\be
k_L=\frac{2\pi}{L}.
\ee
Then since the string thickness $r_0$ is fixed $r_0=1$, the uniform strings are characterized by the value of $k_L$. The smaller values of $k_L$ correspond to the thinner black strings.
\begin{figure}
\begin{centering}
\begin{tabular}{ccc}
\includegraphics[scale=0.3]{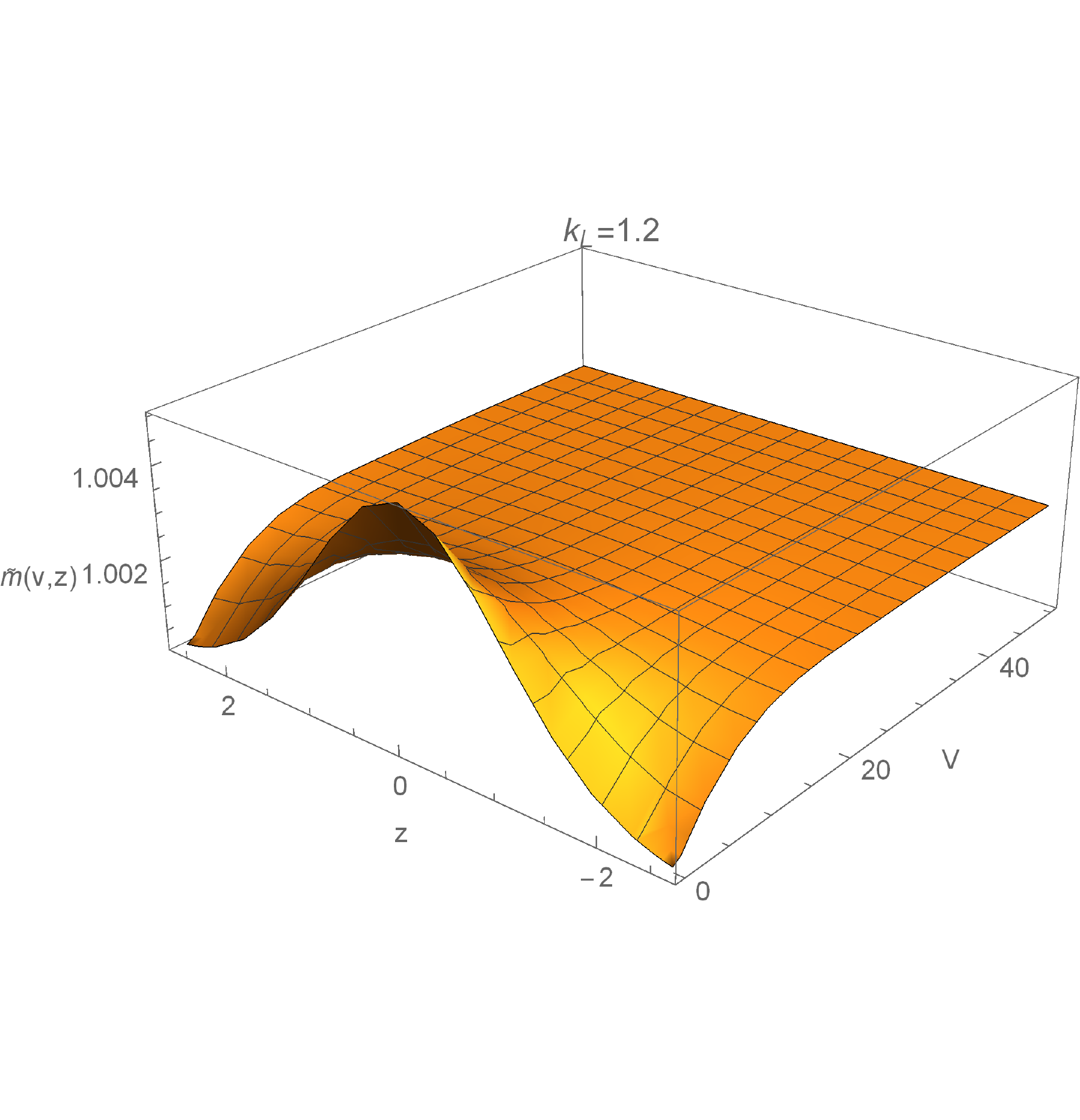} & \includegraphics[scale=0.3]{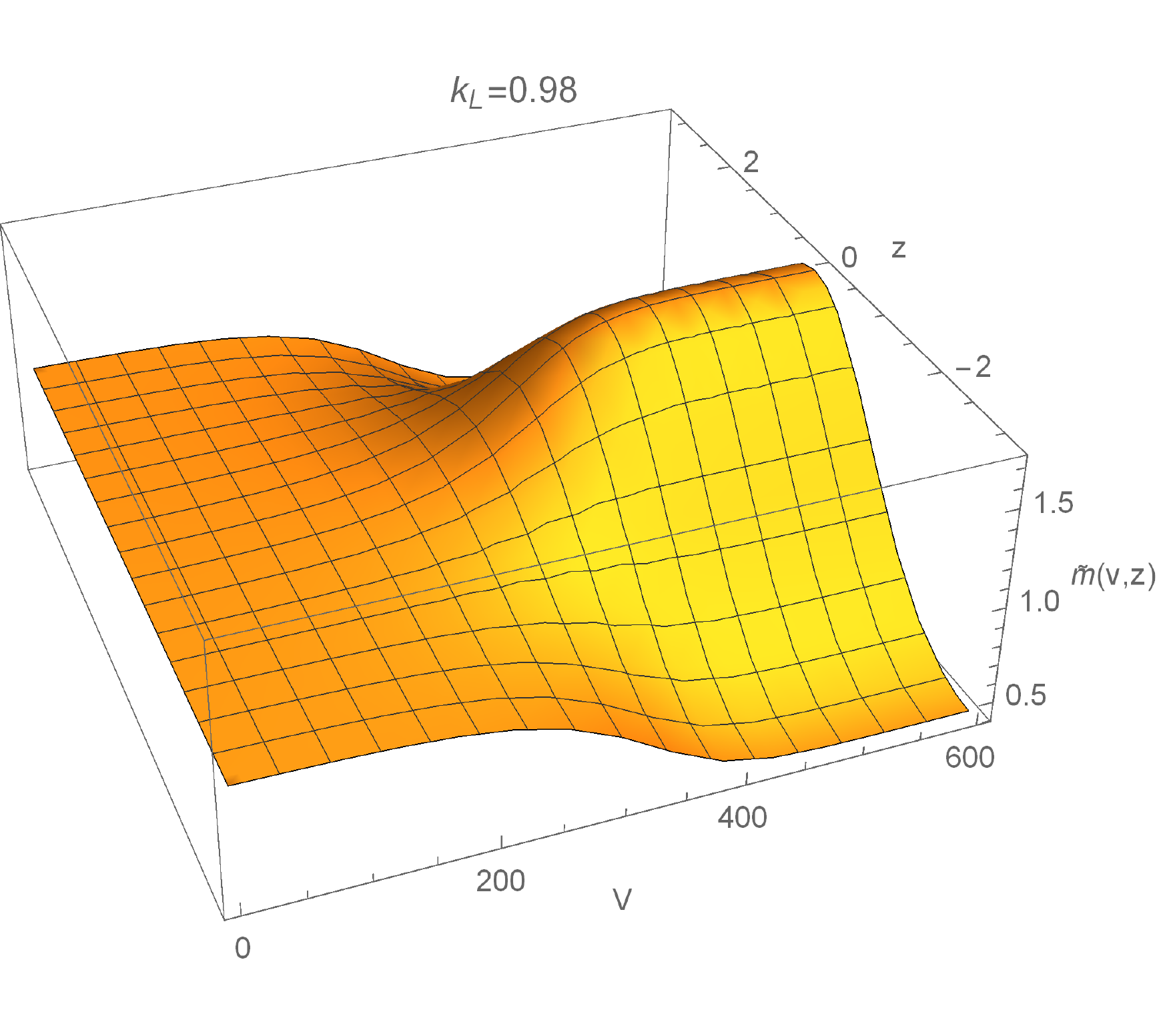} & \includegraphics[scale=0.3]{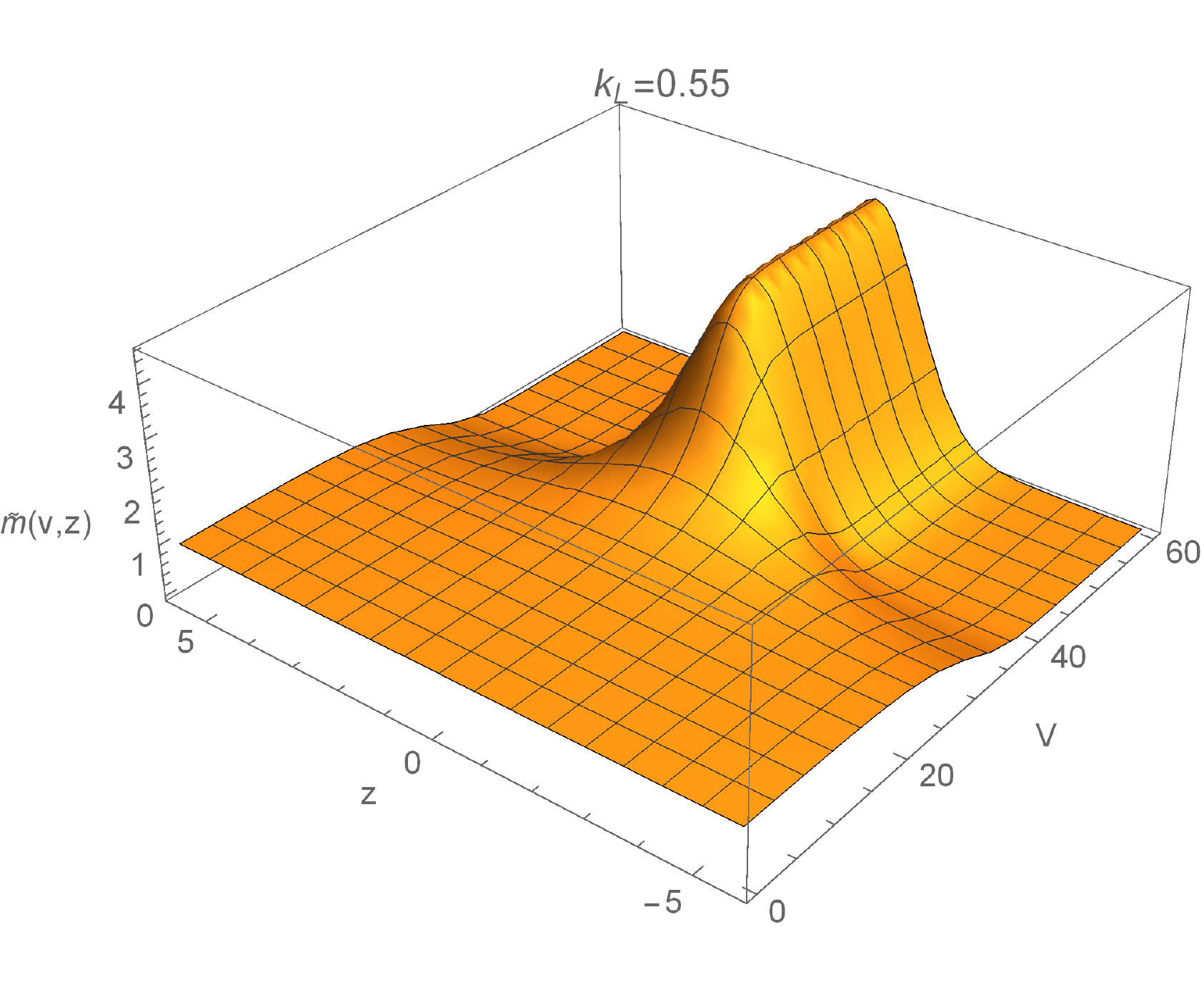}\tabularnewline
\end{tabular}
\par\end{centering}
\caption{\label{fig:UStable}The  dynamical evolution of a perturbed EGB black string with $\tilde{\alpha}=1$. From left to right $k_L=1.2$, $0.98$ and $0.55$, which corresponds to
a fat, a not-too-thin and a thin black string, respectively. }
\label{massvz}
\end{figure}

We fix the value of $k_L$ and introduce a small perturbation of the static uniform black string, $m(0,z)=m_0+\delta m(z)$, $p_z(0,z)=\delta p(z)$. We find that when $k_L>1$, the perturbation quickly dissipates and the black string becomes uniform, this is in accord with the perturbation analysis in section \ref{QNM} that the linear modes are stable with
a wavenumber $k_L>1$. On the contrary, when $k_L<1$, the initial deformation
grows fast, eventually the black string settles down at a stable  state that approximates very well the NUBS obtained as the static solution of the effective equations (\ref{effeq1}) and (\ref{effeq2}). Therefore, the dynamical evolution of the unstable EGB black string is very similar to case in the Einstein gravity \cite{Emparan1506}.
\begin{figure}
\begin{centering}
\begin{tabular}{cc}
\includegraphics[scale=0.45]{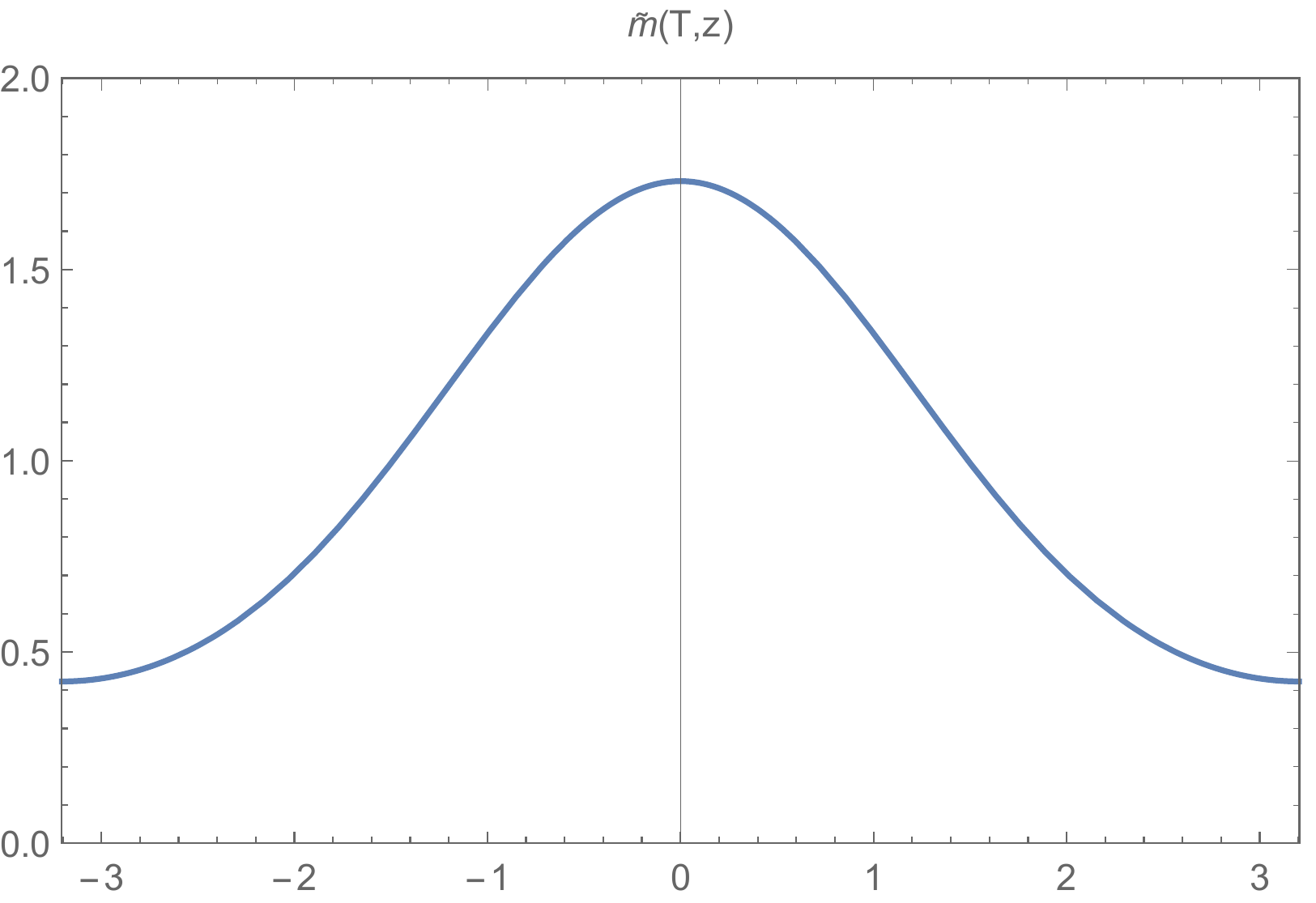} & \includegraphics[scale=0.45]{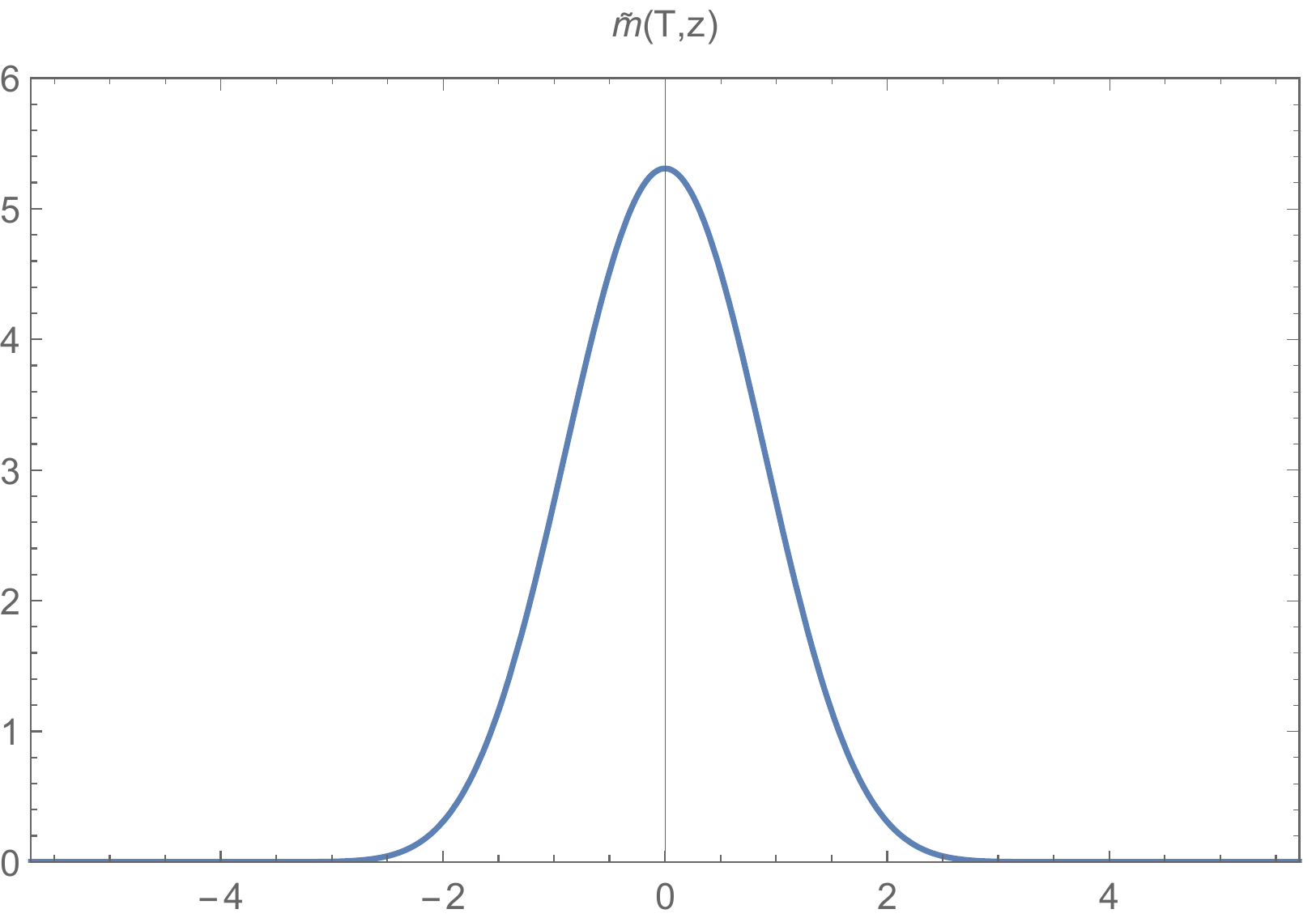}\tabularnewline
\end{tabular}
\par\end{centering}
\caption{\label{fig:55Time12} The  final state of the unstable EGB black strings with $\tilde{\alpha}=1$. The left panel shows the case of a not-too-thin black string with $k_L=0.98$, the right panel shows the case of a thinner black string with $k_L=0.55$. The profile of the left panel is well approximated by a cosine function and the profile of the right panel is well approximated by a gaussian function. }
\label{finalstate}
\end{figure}

In Fig. \ref{massvz} we show the complete evolutions of a fat, a not-too-thin and a thinner black string by plotting $\tilde{m}(v,z)$. The final state of the unstable black strings is shown in Fig. \ref{finalstate}. For a not-too-thin  black string with $k_L=0.98$ the evolution costs much time to reach the final stable state, since $k_L$ is  near $k_{GL}=1$ so the growth rate is small. The final profile is approximately cosinoidal which agrees with discussion of the near uniform solution of the static effective equations (\ref{smalldeformation}). For a thinner black string with $k_L=0.55$, the evolution is faster and its final state has a large blob. In the case of the Einstein gravity, the profile of the large blob is very approximately gaussian, here we find the similar picture  for the EGB black strings.

\begin{figure}
\begin{centering}
\begin{tabular}{cc}
\includegraphics[scale=0.45]{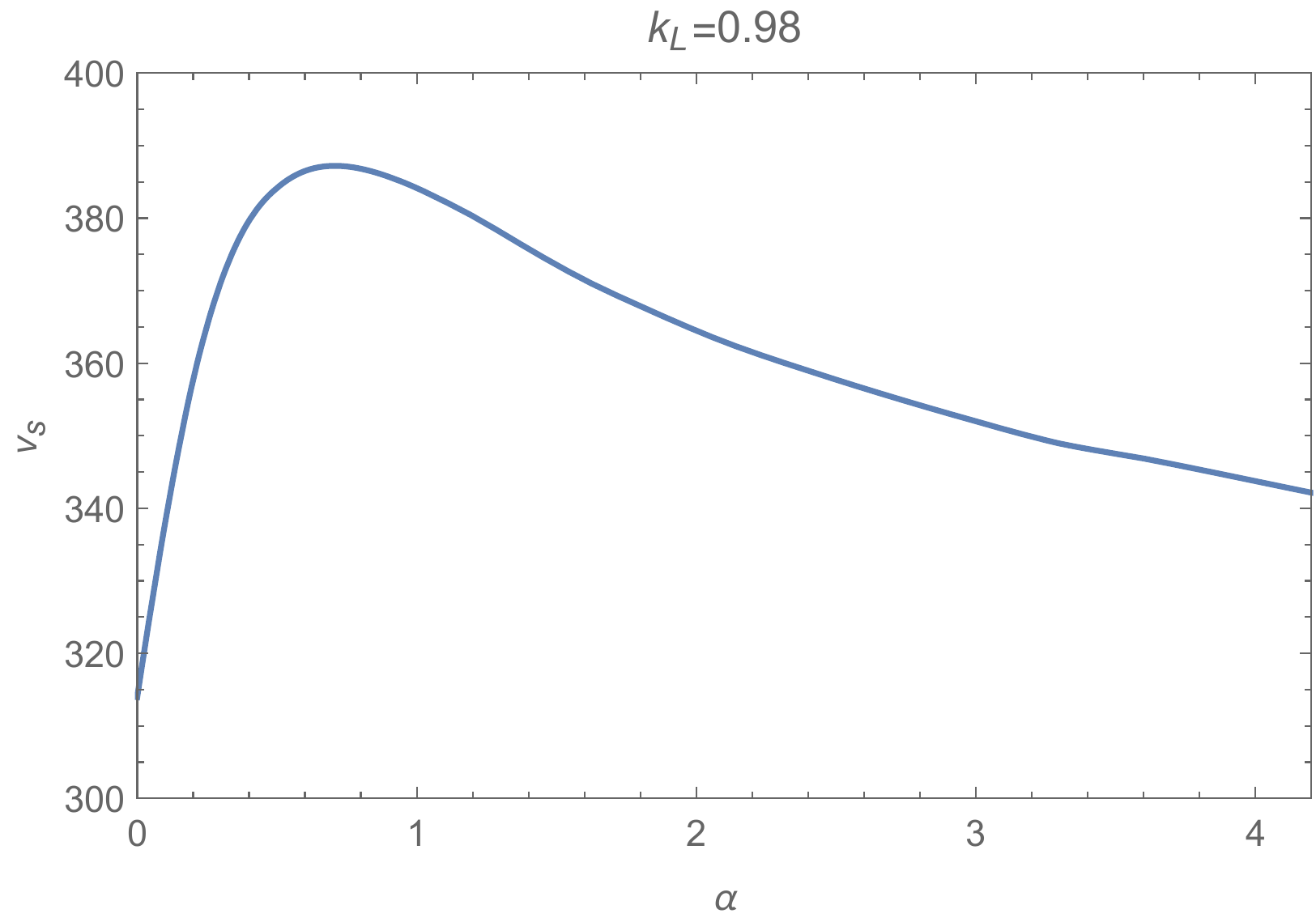} & \includegraphics[scale=0.45]{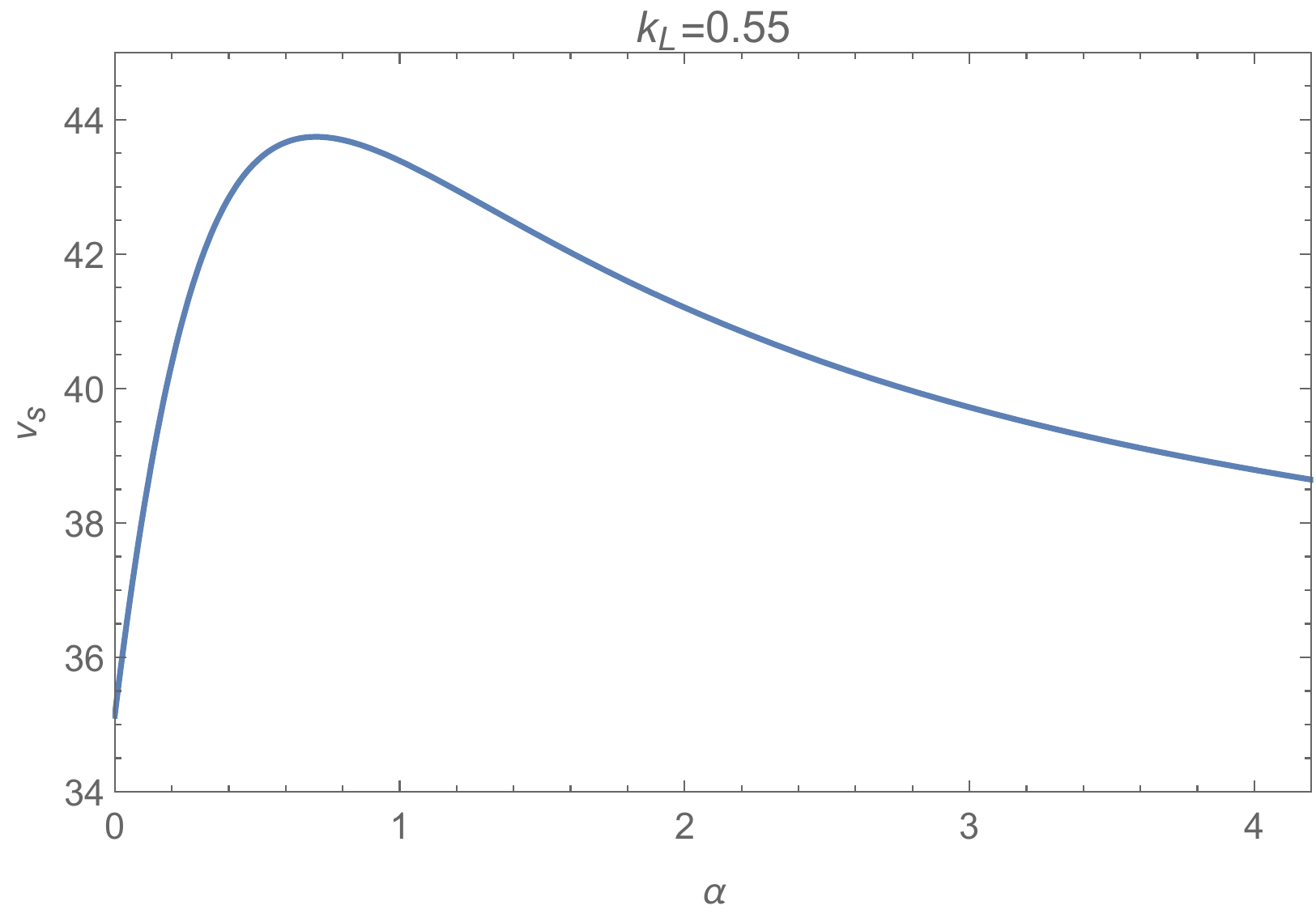}\tabularnewline
\end{tabular}
\par\end{centering}
\caption{\label{fig:55Time}The time $v_{s}$ needed to reach the final state
with respect to different $\tilde{\alpha}$. We see that when $\tilde{\alpha}<0.708$,
the time needed to reach the final state increases with $\tilde{\alpha}$. When $\tilde{\alpha}>0.708$,
the time needed to reach the final state decreases with $\tilde{\alpha}$.}
\end{figure}

According to the perturbation analysis of the effective equations, the GB term affects the growth rate of the unstable modes. Thus we expect that this is reflected by the evolution rate of the black strings. Indeed,  as shown in Fig. \ref{fig:55Time},  when $\tilde{\alpha}<0.708$, the time needed to reach the final state increases with $\tilde{\alpha}$ monotonically, however when $\tilde{\alpha}>0.708$, the time needed to reach the final state decreases with $\tilde{\alpha}$, and when $\tilde{\alpha}$ becomes very large the time to reach the final state approaches to the one in the case $\tilde{\alpha}=0$. We find the turning point is $\tilde{\alpha}\simeq0.708$ for all   $1/2<k_{L}<1$ within the range of numerical validity. (Note that this is close to the QNM turning
point $\tilde{\alpha}=1/\sqrt{2}$.)

\section{Summary}\label{section4}

In this article we studied the black strings in the Einstein-Gauss-Bonnet (EGB) theory by using the large $D$ effective theory. Inspired by the properties of the black strings in the Einstein gravity and the EGB black holes at large $D$, by embedding the effective membrane in the background $\mathcal{M}^{D-1}\times S^1$,  we obtained the effective equations for the EGB black strings after integrating the radial direction of the EGB equations. The EGB black strings were obtained as the static solutions of the effective equations. The uniform black string (UBS) was constructed analytically and it was shown that the leading order solution at large $D$ was identical to a spherically symmetric black hole plus a trivial direction. The non-uniform black strings (NUBS) bifurcating from the UBS at the threshold of the Gregory-Laflamme (GL) instability were studied by numerically solving the effective equations. We found that like the NUBS in the Einstein gravity the profile of large deformations is very approximately gaussian. By performing perturbation analysis of the effective equations we obtained the quasinoramal modes (QNM) of
the EGB black strings. As in the case of the Einstein gravity, the GL instability occurs when the black string is relatively thin. Moreover, we found that there exists a critical value for the GB coefficient, below which the presence of the GB term makes the instability weaker, above which the instability gets enhanced.

Furthermore we also numerically solved the effective equations to study the non-linear evolution of the EGB black strings. The behavior in the non-linear regime is basically similar to the case of the Einstein gravity. For fat black strings the initial perturbation quickly dissipates and the black string becomes uniform which is in accord with the perturbation analysis of the effective equations. For thinner black strings the initial deformation grows fast, finally the black string settles down at a stable NUBS. The effect of the GB term is reflected in the time the unstable black strings needed to reach the final stable state. We found that when $\tilde{\alpha}<0.708$ the time increases, when $\tilde{\alpha}>0.708$ the time decreases, which qualitatively agrees with the effect of the GB term on the GL instability.

On the other hand, we may interpret the large $D$ effective equations for the EGB black strings as the equations for the dynamical fluid. From this point of view, the pressure of the fluid is negative and the speed of the sound of the long-wavelength perturbation is imaginary, which signals  the   GL instability, similar to the case  in the asymptotically flat Einstein gravity. Furthermore we evaluated the transport coefficients and found that the KSS bound for the ratio of shear viscosity and the entropy density ($\eta/s$) is violated for any positive GB coefficient. Our result for  $\eta/s$ is the same as the one obtained in the membrane paradigm in the large $D$ limit \cite{Jacobson1107}.

The work in this paper can be extended in several directions. For example, as discussed in \cite{Suzuki1506} by adding the $1/n^2$ corrections to the effective equations the large $D$ analysis gives a critical dimension $D*\simeq 13.5$ at which the phase transition between the UBS and the NUBS changes from first order to second order. It would be interesting to investigate if the same phenomenon occurs for the EGB black strings and the effect of the GB term on the critical dimension by using the $1/n$ expansion method. Another extension is to consider the black ring solution in the EGB theory \cite{Chen1707}, until now this is basically an unexplored problem. As we mentioned before since the tension of the large $D$ EGB black string is small compared with its mass, such that the rotation of the bent rotating black string used to balance its  tension is small, as a consequence the construction of the corresponding black ring solution is possible. It would be interesting to study the EGB black strings in the framework of the large $D$ membrane paradigm developed
recently as well \cite{Bhattacharyya1504, Bhattacharyya1511, Dandekar1607, Dandekar1609, Bhattacharyya1611, Bhattacharyya1704}. Another interesting subject is on the asymptotically AdS black objects in the EGB gravity at large $D$. This may shed light on the large $D$ limit of the turbulence \cite{Yarom} from holographic point of view.

\section*{Acknowledgements}

 We thank the participants in ``Gravity - New perspectives from strings and higher dimensions" at Centro de Ciencias de Benasque Pedro Pascual for helpful discussions.
BC was in part supported by NSFC Grant No.~11275010, No.~11335012 and No.~11325522. C.-Y.Zhang is supported by National Postdoctoral Program for Innovative Talents BX201600005.

\appendix
\section{The case $\alpha=\mc O(1)$}\label{appendixA}
In this appendix we study the $1/n$ expansion of the EGB black strings in the case  $\alpha=\mc O(1)$. In this case $\tilde{\alpha}=\mc O(n^2)$ which is very large in the large $n$ limit.
From the action (\ref{action}) one can see that now the GB term becomes dominant since $R=\mc O(n^2)$, and the effect of the Einstein gravity is pushed to the third order of
the $1/n$ expansion, so the leading order solutions and the effective equations stem from the pure GB term.

In this case the boundary conditions (\ref{bdycondition}) are not valid anymore. Instead, the proper
ones are given by
\be
A=1+\mc O(\sR^{-1/2}),\quad C_z=\mc O(\sR^{-1/2}),\quad G_{zz}=\frac{1}{n}\Big(1+\mc O(\sR^{-1/2})\Big),
\ee
which come from the discussion of the decoupled quasinormal modes of the EGB black holes\cite{Chen1511}.
The leading order solutions of the EGB equations are
\be
A=1-\sqrt{\frac{\mathbf{m}(v,z)}{\sR}},\quad C_z=\frac{\mathbf{p_z}(v,z)}{\sqrt{\mathbf{m}(v,z)}\sqrt{\sR}}
\ee
\be
u_v=1,\quad G_{zz}=\frac{1}{n}\Big(1+\frac{1}{n}\frac{\mathbf{p_z}(v,z)^2}{\mathbf{m}(v,z)^2\sqrt{\sR}}\Big),
\ee
where $\mathbf{m}(v,z)$ and $\mathbf{p_z}(v,z)$ are introduced as the integration functions of the $\sR$-integrations.
The above formulae are related to (\ref{LOAm}) and (\ref{LOCzGzz}) by the following relations
\be\label{relations}
m(v,z)=\tilde{\alpha}\,\mathbf{m}(v,z),\quad p_z(v,z)=\tilde{\alpha}\, \mathbf{p_z}(v,z),
\ee
when taking the large $n$ limit.

The effective equations are
\be
\partial_v\mathbf{m}-\partial_z^2\mathbf{m}+ \partial_z\mathbf{p_z}
=0,
\ee
\be
\partial_v \mathbf{p_z}-\partial_z^2 \mathbf{p_z}-\partial_z\mathbf{m}+\partial_z \Big(\frac{\mathbf{p_z}^2}{\mathbf{m}}\Big)=0.
\ee
They can also be reproduced from  (\ref{effeq1}) and (\ref{effeq2}) by using the relations (\ref{relations}). It is easy to find that the effective equations are exactly the
same as the ones in the Einstein gravity at large $D$. Therefore the assumptions for the large $D$ scaling of the GL threshold modes and the metric
components are self-consistent.

\end{document}